# High magnetic field evolution of the in-plane angular magnetoresistance of electron-doped $Sr_{1-x}La_xCuO_2$ in the normal state

V. P. Jovanović,[1] H. Raffy,[2] Z.Z. Li,[2] G. Reményi,[3] and P. Monceau[3,4]

[1]University of Belgrade, Institute for Multidisciplinary Research, Kneza Višeslava 1, 11030 Belgrade, Serbia.
[2]Laboratoire de Physique des Solides, CNRS UMR 8502, Bâtiment 510, Université Paris-Sud, Université Paris-Saclay, 91405 Orsay, France.
[3]Université Grenoble Alpes, CNRS, Grenoble INP, Institut Néel, 38000 Grenoble, France
[4]Université Grenoble Alpes, INSA Toulouse, Université Toulouse Paul Sabatier, CNRS LNCMI, 38000 Grenoble, France.

We studied the in-plane angular magnetoresistance (AMR), in the normal state, of underdoped superconducting $Sr_{1-x}La_xCuO_2$, which has the simplest crystal structure among cuprates. The measurements of two underdoped thin films with different dopings were performed in intense magnetic field $H$ (up to 22 T). The longitudinal magnetoresistance at temperature $T$ is negative and scales with $H/T$. For both samples, the AMR is anisotropic and shows an unexpected dependence on $H$ intensity. While at the low magnetic field, one observes essentially twofold AMR oscillations for the more doped sample, fourfold ones start to grow under the high magnetic field, resulting in the coexistence of the two. For the less doped film at the low magnetic field, both twofold and fourfold AMR components exist. With the increase of the magnetic field, the fourfold component survives a $\pi/4$ phase shift, during which its amplitude vanishes, at a magnetic field $H_c$ such as: 16 T < $H_c$ < 17 T. As a result, at the high magnetic field above $H_c$, the angular dependence of the in-plane magnetoresistance turns out to be the same for both samples. We tentatively ascribe the above features to the presence of anti-ferromagnetism in the $CuO_2$ planes of underdoped $Sr_{1-x}La_xCuO_2$.

## I. INTRODUCTION

One important and intriguing question in the study of high-transition temperature (high-$T_c$) cuprates is the role of magnetism in their transport and superconductivity properties. It is well known that in cuprates, superconductivity emerges by doping with holes (*h*) or electrons (*e*) from an insulating antiferromagnetic (AF) phase. In *h*-doped cuprates, AF is suppressed rapidly upon the establishment of the superconductivity. In the case of *e*-doped cuprates, the phase diagram, temperature $T$ vs. *e* doping, appears to be very different (see Fig. 2 of [1]), one of the main differences being the extent of the AF phase, which persists to much higher doping than in *h*-doped cuprates. The AF phase appears to be much more robust and coexists with superconductivity in the underdoped side of the phase diagram. One possible explanation is a dilution of the spin system by neutralization of the spin on the $d^9$ Cu site with *e* doping, in a way similar to the effect of $Zn^{2+}/Cu^{2+}$ substitution [1]. In contrast, for *h* doping, the doped holes primarily reside on in-plane oxygen and produce a spin frustration with a rapid decrease of AF phase. This magnetic property has been the focus of the investigation in the most studied *e*-doped family $Ln_{2-x}Ce_xCuO_4$ where *Ln* is lanthanide rare earth: Pr, Nd, Sm, and La [1]. As single crystals are available for these compounds (except for $La_{2-x}Ce_xCuO_4$), the AF structure has been observed in neutron scattering experiments [1], the ideal technique to obtain the spin arrangement up to high magnetic fields. There are also local probes such as nuclear magnetic resonance (NMR) and muon spin spectroscopy (μSR). The first one requires bulk samples, while the latter is operated in relatively low magnetic fields.

There exists another class of *e*-doped cuprates, $Sr_{1-x}La_xCuO_2$ (SLCO), which is much less studied. For technical reasons, the synthesis of single-crystal has not been possible, requiring very high pressure, and thus only thin films have been prepared. Still, SLCO exhibits several favorable characteristics. The most interesting one is its crystal structure, which is the simplest of all the cuprates. It consists (see inset of Fig. 1) of a stack of flat planes of square $CuO_2$ lattice separated only by Sr atoms partially substituted by La for *e* doping ($La^{3+}/Sr^{2+}$). The doping can also be achieved by removing the oxygen atoms localized between $CuO_2$ planes during the synthesis. Compared to the $Ln_{2-x}Ce_xCuO_4$ family, there is no magnetic atom in the structure which has been shown to influence the magnetism in the $CuO_2$ planes, the adjacent $CuO_2$ planes are identical [no shift (a/2, a/2) with respect to each other], and the interplane distance is shorter: 3.4 Å instead of 6 Å. SLCO belongs to the family of the so-called infinite layer (IL) compounds. Interest for this phase has been renewed by the recent discovery of another IL *h*-doped superconductor $Nd_{1-x}Sr_xNiO_2$ [2] with $T_c$ = 15 K. As both compounds have the same crystal structure, it will be interesting to compare in the future their phase diagrams and electronic properties [3,4].

The fundamental question that is raised in this paper is the existence of AF in SLCO. Due to the absence of single-crystal sample, neutron experiments have been only made on polycrystalline samples [5], as well as muons experiments in low magnetic fields [6] to study the London penetration depth. Importantly, an angle-resolved photoemission spectroscopy (ARPES) experiment conducted on superconducting epitaxial SLCO thin films (x = 0.1, $T_c$ = (25 ± 5) K [7]) pointed to the existence of a robust $(\pi,\pi)$ AF in SLCO, implying unusually strong coupling of electrons to AF. To complement dedicated facility based measurements, one can use the angular magnetoresistance (AMR), measured by rotating the magnetic field within the $CuO_2$ plane, to probe the spin structure of a conducting sample through the electron-spin interaction. It allows the observation of the AMR anisotropy (with twofold or fourfold symmetry) for several different transition-metal systems, including high-$T_c$ superconductors (both $e$ and $h$ doped) [8-14], ferromagnetic samples such as manganite systems $La_{2/3}Ca_{1/3}MnO_3$ and $La_{2/3}Sr_{1/3}MnO_3$ [15-17], magnetite $Fe_3O_4$ [18-20], and even (Ga, Mn)As [21] (which is not an oxide and whose properties are dependent on $d$ electrons of Mn). Very recently, AMR with a fourfold symmetric component above a critical magnetic field has been reported in strontium iridate $SrIrO_3$ thin films to probe the correlations between electronic transport and magnetic order [22]. For cuprates, it has been suggested that the origin of AMR is due to long-range AF [11], but also spin-charge segregation or stripes [8-10], spin-orbit induced anisotropy [23], and spin flop in cuprates with a magnetic ion [12, 24]. In the other, non-cuprate materials we mentioned, spin-orbit induced origin [15, 16, 18, 21], the formation of antiphase boundaries in samples [19], and the influence of lattice symmetry [25] were proposed. Although there are various explanations of this phenomenon, one may still wonder if this fourfold-symmetric AMR may have a common origin for all of these materials.

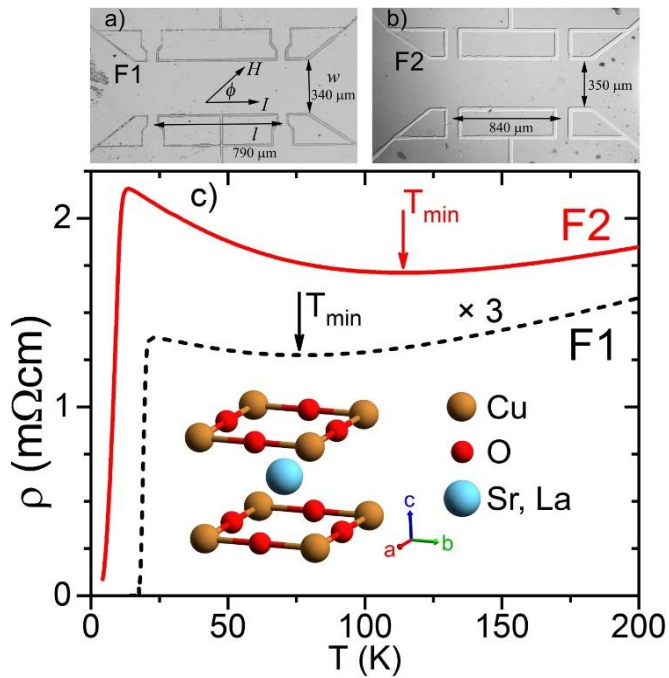

FIG. 1. Pictures of thin films (a) F1 and (b) F2, patterned by optical lithography. (c) Resistivity $\rho$ of films F1 (dashed black curve, multiplied by 3) and F2 (solid red curve). Arrows indicate resistivity minima $T_{min}$. Inset shows the crystal structure of SLCO films.

In our previous work (Ref.[14]) four underdoped, c-axis oriented, epitaxial SLCO thin films (named 1-4, with increasing $T_c$ [1≤$T_c(\rho=0)$≤17K]), were studied in the normal state (T < 75K) under a magnetic field up to 6T. Summarizing briefly the results it was shown that the in-plane magnetoresistance is negative, does not saturate with increasing field, and has a doping-dependent anisotropy: a twofold AMR was found for the more doped samples (3,4) and a superposition of twofold and fourfold AMR for the less doped samples (1,2), both increasing with increasing magnetic field and decreasing temperature. It was concluded that the most probable origin of fourfold oscillations was the presence of AF. The present paper reports the study of magnetoresistance (MR) and AMR of two of these films (2 and 4) up to 22 T (Grenoble High Magnetic Field Laboratory). It describes how the two typical AMR situations, twofold and fourfold, evolve under the increase of the applied magnetic field, and how this can be related to the presence of AF. It will be shown that the magnetic field strongly influences the spin configuration, as manifested by a field-induced transition with $\pi/4$ change in the phase of the fourfold oscillations (or alternatively as a change of sign of the amplitude of fourfold oscillations crossing zero) at a field between 16 and 17 T for the weakly doped sample and the progressive growth of fourfold oscillations for the more doped one.

## II. EXPERIMENTAL TECHNIQUES

The synthesis of the $Sr_{1-x}La_xCuO_2$ (x = 0.12) thin films of this paper, labeled F1 and F2, was previously described in Ref. [26]. Briefly, they were epitaxially grown by single-target, rf magnetron sputtering on $KTaO_3$ single-crystal (001) substrates and in-situ reduced during the cooling stage of the synthesis. These films are single-phase, c-axis oriented as confirmed by x-ray diffraction spectra. Their thicknesses are 90 nm for F1 and 60 nm for F2.

The two samples were previously studied in a low magnetic field [14]: $H \leq 6$ T, F1, and F2 corresponding respectively, to films 4 and 2 of Ref. [14]. They are both underdoped, F1 being more $e$ doped (high doping) than F2 (low doping). The different doping states were obtained by different in situ oxygen reduction after each film deposition, as explained in Ref. [26].

The reported magnetotransport measurements were performed in LCMI-Grenoble in a 22-T Bitter coil. Both samples were glued on the rotating platform (horizontal axis) of the sample holder, allowing the magnetic field to rotate in the $(a, b)$ plane of the samples. The resistance of the Hall bar patterned films [see Figs. 1(a) and 1(b)] was measured with a four-probe ac method in the temperature range 15 K $\leq T \leq$ 50 K with the current (typically 250 µA) flowing along the $a$-axis of the films. The temperature was regulated with a capacitive sensor while the temperature in zero magnetic field was measured with a Cernox thermometer. Measurements were conducted of the normal-state MR, $\Delta\rho(H, T) = \rho(H, T) - \rho(0, T)$, at given T and the angular dependence of the MR, $\Delta\rho(\phi, H, T)$, for given T and H, where $\phi$ is the angle between the film $a$ axis and the magnetic field $H$.

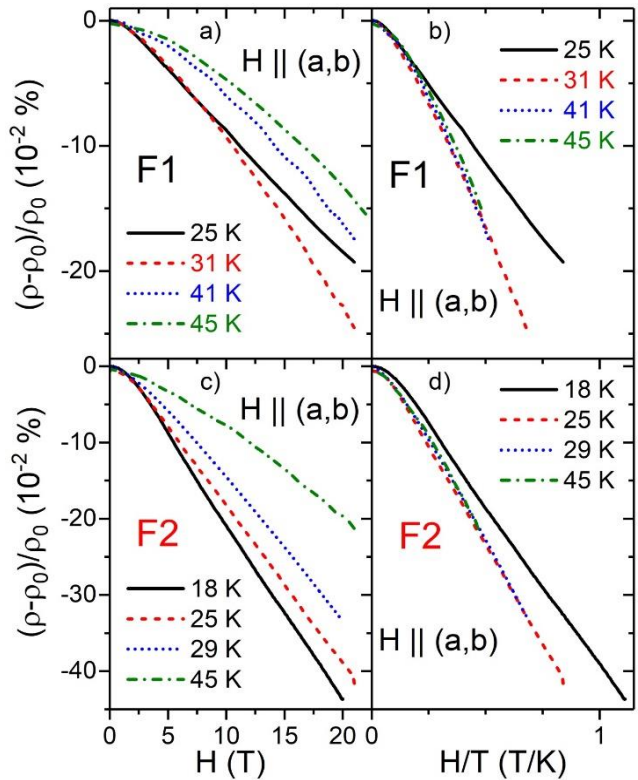

FIG. 2. (a) Normalized magnetoresistance $(\rho - \rho_0)/\rho_0$ of F1 for magnetic field $H$ parallel to $CuO_2$ planes at different temperatures. (b) The same normalized MR as a function of $H/T$. (c) Normalized MR of F2 for magnetic field $H$ parallel to $CuO_2$ planes at different temperatures. (d) The same normalized MR as a function of $H/T$. The $\rho_0$ is the zero-field resistivity, curves are smoothed, and measurement error is estimated to $10^{-3}$ %.

## III. EXPERIMENTAL RESULTS

### A. Low-temperature resistance and in-plane longitudinal magnetoresistance

Figure 1(c) shows the resistive superconducting transition of films F1 and F2. For F1, the more *e*-doped sample, $T_c$ ($R = 0$) is 17 K, and for F2, the less doped sample, $T_c$ ($R = 0$) is less than 4 K. As shown earlier [27], there is an upturn of the low-temperature resistivity below a temperature $T_{min} = 77$ K for F1 and $T_{min} = 114$ K for F2. It can be seen that the less doped sample F2 exhibits a higher normal-state resistivity, a larger low-$T$ resistance increase, and a lower $T_c$ than the more doped sample F1.

Shown in Fig. 2 is the in-plane magnetoresistance $\Delta\rho(H, T)$ of F1 and F2 measured at given temperatures $T$ in a magnetic field up to 22 T parallel to the *a*-axis of the films. The magnetoresistance is negative, its absolute value increasing linearly without saturation with increasing field $H$ and decreasing with increasing $T$. In Ref. [14], it was found that it disappears for $T$ around 60 K for $H = 6$ T. By plotting these results versus $H/T$ in Figs. 2(b) and 2(d), one observes a scaling of the magnetoresistance except for the curves measured at the lowest temperature, i. e. closest to the onset of superconducting fluctuations, the latter giving a positive contribution to the MR.

### B. Angular dependence of the in-plane magnetoresistance

Shown in Fig. 3 is the polar representation of the angular dependences of MR (AMR) measured at different constant magnetic fields rotating within $CuO_2$ planes: 10, 15, and 21 T [Figs. 3(a)-3(c), respectively] for F1 (high doping) and 10, 16, and 21 T [Figs. 3(d)-3f), respectively] for F2 (low doping). The angle $\phi = 0$ corresponds to $H$ parallel to the *a* axis (the inset in Fig. 4). Both films were measured simultaneously on the same rotating sample holder which allows a direct comparison between the different magnetoresistance behavior of both samples and which precludes the effect of possible misalignment. At *low* magnetic fields ($H \leq 14$ T), fourfold angular oscillations of MR are only observed at low doping (film F2), while at high doping (film F1), only twofold oscillations are visible. This result obtained in rotating the samples in a fixed magnetic field is in agreement with our previously reported measurements This precludes also the effect of possible experimental rotation artifacts on the results. The measurements performed under a higher magnetic field unveil the observation of a dramatic change of behavior of the less doped sample F2 [compare Figs. 3(d) and 3(e)) leading both samples F1 and F2 to exhibit the *same* behavior at high magnetic fields [compare Figs. 3(c) and 3(f)]. Such a behavior constitutes the central result of this paper, which will be discussed in the following.

## IV. DISCUSSION OF THE RESULTS

### A. Spin origin of the longitudinal magnetoresistance

A negative longitudinal magnetoresistance (MR) was already reported by different authors in various underdoped cuprates [13,28-30]. In underdoped *e*-type $Nd_{2-x}Ce_xCuO_4$ (NCCO), Tanda *et al*. [28] attributed this MR behavior to localized spins. For very underdoped (x = 0.1), *h*-type $La_{2-x}Sr_xCuO_4$ (LSCO) Preyer *et al*. [29] also showed a negative MR linear with $H$ that scales with $H/T$ only in the low-$T$ resistance upturn region. They assigned the MR scaling as a function of $H/T$ to a reduction of spin scattering by the magnetic field, resulting from a strong coupling between the fluctuating Cu spins and the charge carriers. In *e*-doped $La_{2-x}Ce_xCuO_4$ superconducting thin films, Jin *et al*. [13] reported a negative in-plane MR only for low doping, less than 0.08, where MR turned to be positive for higher x, while Preyer *et al*. [29] observed that the MR in the normal state of $La_{2-x}Sr_xCuO_4$ is negative and independent of the orientation of the magnetic field parallel or perpendicular to the $CuO_2$ planes. We note that in the normal state of our SLCO thin films the MR in a magnetic field perpendicular to the $CuO_2$ planes is always positive and much larger than the longitudinal one [30] and essentially due to superconducting fluctuations.

That the negative MR could be due to the Kondo effect in heavily underdoped *non-superconducting* cuprates ($A_{2-x}Ce_xCuO_4$, $A$ = La, Nd, Pr) was proposed by Sekitani *et al*. [31]. They found a negative MR, with log$T$ behavior (followed by saturation at low $T$) of the $R(T)$ upturn.

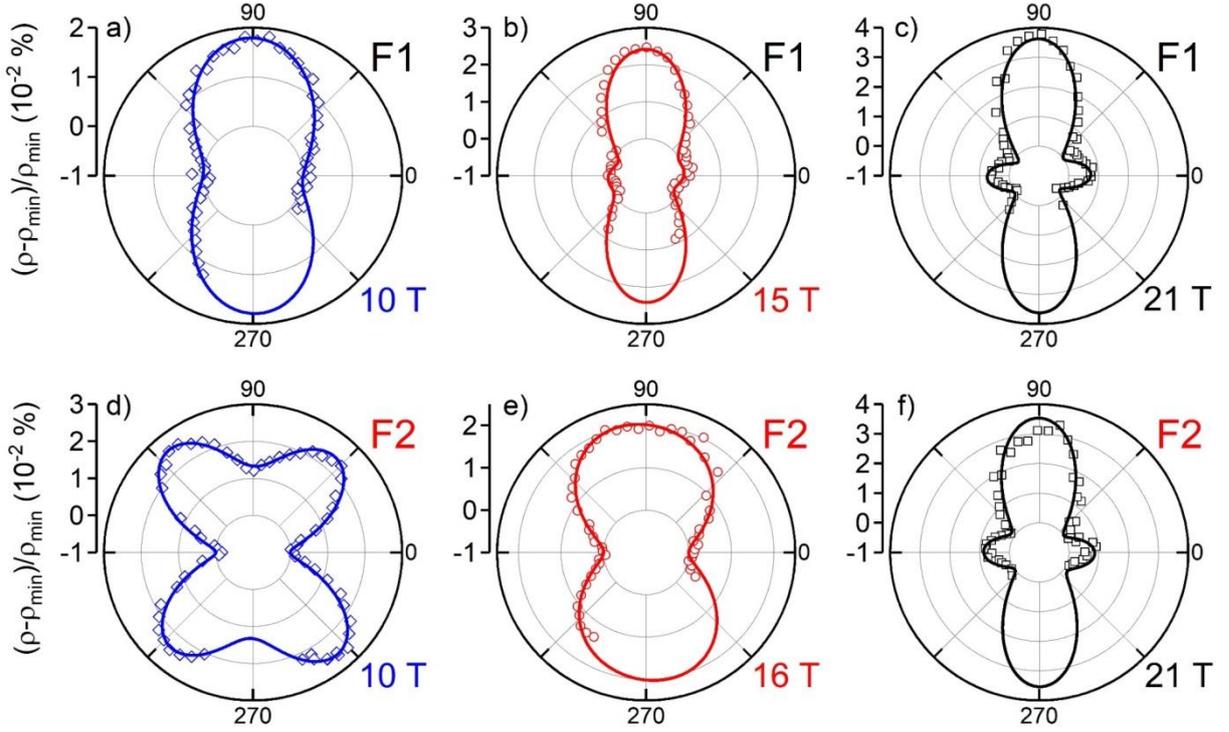

FIG. 3. Polar plots of AMR of F1 (high doping) at 25 K are given in (a) for 10 T, (b) for 15 T, and (c) for 21 T. Polar plots of AMR of F2 (low doping) at 18 K are given in (d) for 10 T, (e) for 16 T, and (f) for 21 T. Solid lines represent fits to experimental data given by Eq.(1) while $\rho_{min}$ is the resistivity corresponding to the minimum of AMR curves. The error could be estimated as the average difference between raw data and the fitted curve

The Kondo scattering, which usually produces $R \sim -\log H$, is questionable in our case by linear $H$ dependence of MR and $H/T$ scaling. Moreover, there is no magnetic ion in SLCO. The observed negative MR appears to result from a reduction of spin scattering by the magnetic field. However, following the procedure of Sekitani et al. [31] to determine the Kondo temperature $T_K$ from the resistivity $\rho(T)$ and using the expression given in Ref. [32], we were not able to get physical values of $T_K$ for our films, thus excluding the possibility of $H/(T+T_K)$ scaling of the MR.

Several works have reported on the intrinsic static or quasistatic AF order in e-doped cuprates: in $Pr_{2-x}Ce_xCuO_4$ up to x = 0.16 [11] and in $La_{2-x}Ce_xCuO_4$ [13]. Moreover, the ARPES study on SLCO thin films [7] has reported on the presence of strong AF in SLCO, so it seems reasonable to attribute the behavior of the MR in SLCO to spin scattering in presence of AF. Besides, following the suggestion of Dagan et al. [33] for e-doped $Pr_{2-x}Ce_xCuO_4$ thin films, the spin scattering assumption could also explain the resistivity upturn seen in Fig. 1, and the linear infield magnetoresistance [34].

Contrasting our SLCO are the in-plane MR results for non-superconducting single crystals of e-doped $Pr_{1.85}Ce_{0.15}CuO_4$ [8] and $Pr_{1.3-x}La_{0.7}Ce_xCuO_4$ [12]. In these samples, the MR measured with the field along Cu-O-Cu direction is positive and exhibits a change of slope at $H = 5$ T, more pronounced when measured with the magnetic field along the Cu-Cu axis. This dramatic behavior is due to the spin-flop effect confirmed by neutron scattering experiments (see p. 2436 of [1]). In these compounds, the spin orientation is turned by 90° in adjacent $CuO_2$ layers, exhibiting a noncollinear magnetic structure. Spin-flop transition results in the alignment of the spins in the same direction in all the planes. Although the spin orientation in AF SLCO was not studied experimentally, we may suppose that it is the same as in other AF compounds with the same IL structure. Neutron powder diffraction studies on two IL compounds, $(Ca_{0.85}Sr_{0.15})CuO_2$ (CSCO) [5] and $SrFeO_2$ [35], implies three-dimensional (3D) $(\pi, \pi, \pi)$ antiferromagnetic order with magnetic moments perpendicular to the c-axis. We may suppose that AF SLCO has the same spin configuration (see Fig. 2(b) of Ref. [35]), further evidence for the absence of such a spin-flop transition in our MR curves.

### B. Behavior of the AMR

#### 1. Angular dependence of the magnetoresistance

We will now focus on the anisotropic AMR shown in Fig. 3. Comparing the order of magnitude of the measured AMR (Fig. 3) with that of the longitudinal MR (Fig. 2), the latter is 10 – 20 times larger. Consequently, we may suppose that the MR is composed of two parts, a large isotropic (at least isotropic in $CuO_2$ planes) one, and a small anisotropic one. The isotropic part, resulting in negative MR, is a consequence of the AF background in the $CuO_2$ planes of our underdoped films. The origin of the small anisotropic part of MR is less straightforward. It reveals an anisotropic spin-electron scattering that could correspond to easy axes for spin orientation. For cuprates, one can find in the literature various explanations of anisotropic AMR, like an onset of spin-charge stripes [8,9] or a long-range AF ordering [11, 13]. We may rule out the possibility of the formation of stripes in SLCO primarily because there is no hysteretic behavior of longitudinal MR (Fig. 2) unless the whole sample behaves as a single domain. Other transition-metal oxides, such as the half-metals magnetite and manganites (with ferro- or ferrimagnetic properties), also display AMR anisotropy. For these materials, AMR oscillations have similar origins: Twofold AMR is due to the uniaxial symmetry of the transport current, and the fourfold AMR is assumed to be a consequence of the local, spin-orbit induced orbital deformation [15,17]. In these compounds, rotation of the magnetization changes the orbital overlap between neighboring Mn $d$ and oxygen $p$ electrons, which modulates the conductivity. We may suppose that a similar effect could happen in the case of our SLCO in the normal state.

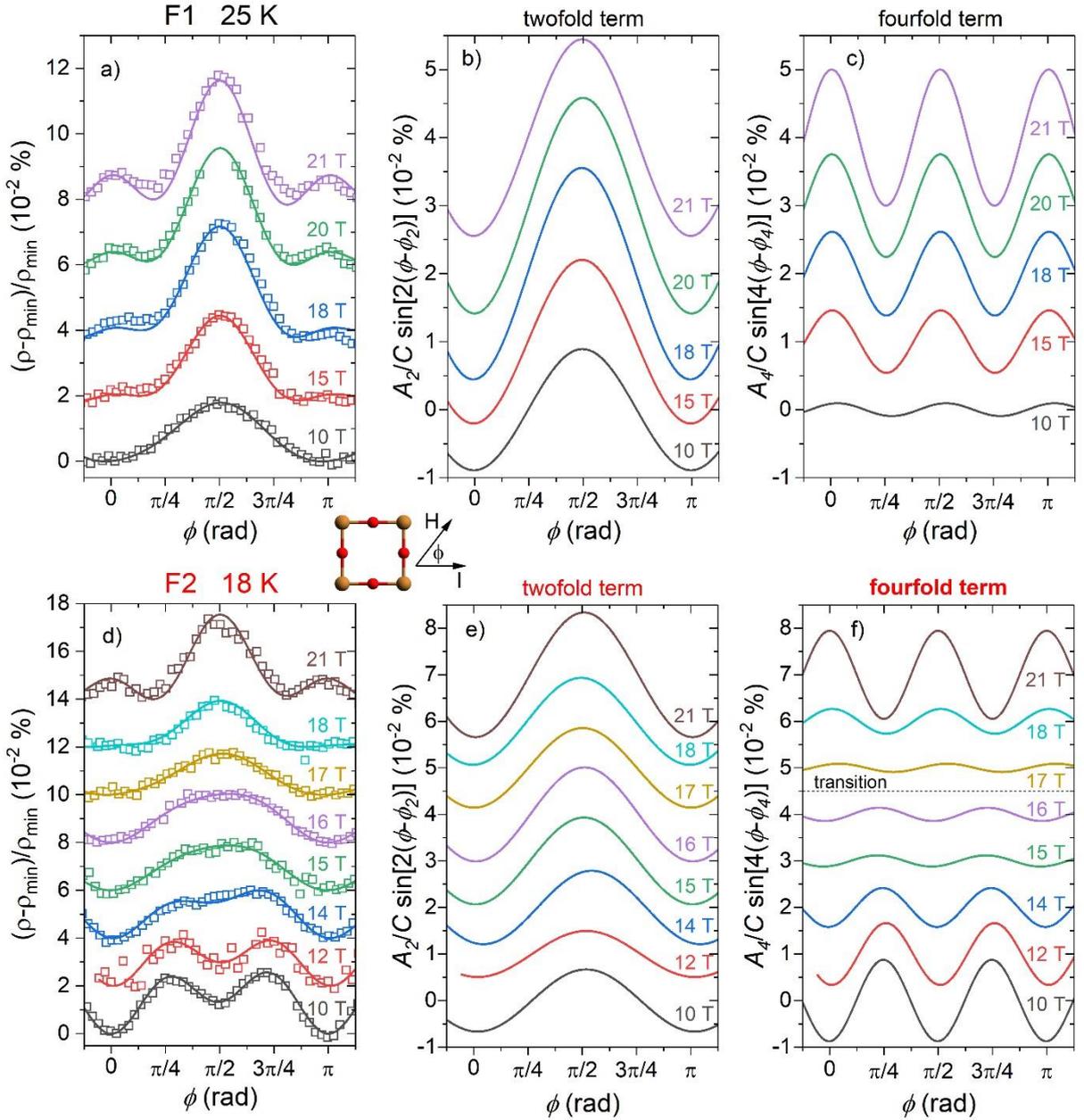

FIG. 4. (a) AMR for film F1 (high doping) at 25 K and 10 T, 15 T (offset by 0.02 %), 18 T (offset by 0.04 %), 20 T (offset by 0.06 %) and 21 T (offset by 0.08 %); (b) twofold and (c) fourfold components of the AMR from Eq. (1) [offsets equal to the half of the corresponding ones from (a)]. (d) AMR for film F2 (low doping) at 18 K at 10 T, 12 T (offset by 0.02 %), 14 T (offset by 0.04 %), 15 T (offset by 0.06 %), 16 T (offset by 0.08 %), 17 T (offset by 0.1 %), 18 T (offset by 0.12 %) and 21 T (offset by 0.14 %); (e) twofold and (f) fourfold components of the AMR from Eq. (1) [offsets equal to the half of the corresponding ones from (d)]. Note that the amplitude $A_4/C$ of the fourfold component crosses zero, or equivalently, the phase of the fourfold oscillation shifts by $\pi/4$ at the transition (dashed) line at a field $H_c$ between 16 and 17 T. The resistivity $\rho_{min}$ corresponds to the minimum of AMR curves. Inset shows the angle $\phi$ between the applied current (along $a$ or $b$) and $H$.

## 2. Field-induced phase shift transition

To analyze in more details our experimental AMR observations, the superimposition of twofold and fourfold oscillations can be fitted, following McGuire and Potter [36], by the following expression:

$$\rho(\phi) = C + A_2 \sin[2(\phi - \phi_2)] + A_4 \sin[4(\phi - \phi_4)], \quad (1)$$

where $C$ is a constant (at given $T$ and $H$); $A_2$ and $A_4$ are the amplitudes of twofold and fourfold oscillations, respectively; and $\phi_2$ and $\phi_4$ are their corresponding phases. We notice that an expression similar to Eq. (1) has been previously derived from a phenomenological model, widely applied to describe the AMR in ferromagnetic thin films (resistivity tensor depending on the direction cosines of the magnetization [17,25]). The fits corresponding to Eq. (1) are drawn as solid lines in Figs. 4(a) and 4(d).

Considering first F1 at the lowest temperature ($T = 25$ K), the more $e$-doped sample, the normalized amplitude $A_2/C$ of the twofold oscillations increases steadily while phase $\phi_2$ remains constant, equal to $\pi/4$ [Figs. 3(a)-3(c) and Figs. 4 (a) and 4(b)]. At high magnetic field above 12 T, a fourfold component starts to appear progressively with amplitude $A_4/C$ much smaller than $A_2/C$ and increasing monotonously with $H$ [Figs. 3(a)-3(c), 4(a), and 4(c)]. The associated phase $\phi_4$ stays constant, equal to $-\pi/8$ in the whole field range [Fig. 4(c)].

Let us consider now the case of the less doped sample F2 at the lowest $T$ ($T = 18$ K). The behavior of the twofold-component parameters $A_2/C$ and $\phi_2$ are similar to those of F1: monotonous increase for $A_2$ and constant value $\pi/4$ for $\phi_2$ [Figs. 4(d) and 4(e)]. Unexpectedly, a dramatic change of behavior occurs for $A_4$ and $\phi_4$ at a magnetic field $H_c$ between 16 and 17 T. The phase $\phi_4$ undergoes a sharp transition from $\pi/8$ to $-\pi/8$ while the amplitude $A_4$ decreases to zero [Figs. 3(e) and 4(f)], followed by a new increase with increasing field [Figs. 3(f) and 4(f)]. Above 18 T it appears that both films F1 and F2 present the same AMR behavior [see Figs. 3(c) and 3(f)]. AMR measurements performed for F2 at higher temperature ($T = 25$ K) at 6, 10 and 21 T have revealed the same shift of $\phi_4$ between low and high magnetic field.

Both films show similar AMR behavior at higher temperatures. At 21 T (film F1, Fig. 5), the oscillations of AMR keep the same shape with increasing temperature, but decreasing amplitude [Fig. 5(b)]. According to Ref. [14] and Fig. 5(b) oscillations of AMR should vanish around 60 K.

## 3. Possible origin of the AMR evolution under high magnetic field

In a previous study under lower magnetic field of the AMR of these SLCO thin films [14] we already discussed the possible origin of the anisotropy. In our discussion, we first ruled out several nonmagnetic origins of the AMR such as the symmetry of the order parameter and the influence of a pseudo-gap with $d$-wave symmetry. We eliminated also the Lorentz force as a possible origin of the twofold oscillations, which decrease when $T$ increases. After examining different mechanisms, we concluded that the most plausible cause of the observed AMR was due to the presence of AF in the CuO$_2$ planes. We propose that the AF still explains the high-field AMR of SLCO. We think it is valid for both the fourfold and the twofold oscillations as concluded also in Ref. [13] where only twofold oscillations were reported. This hypothesis has been strengthened by the ARPES study of SLCO thin films [7] with $T_c = 25$ K, even more $e$ doped than F1, which has revealed a strong $(\pi,\pi)$ AF influence in SLCO thin films.

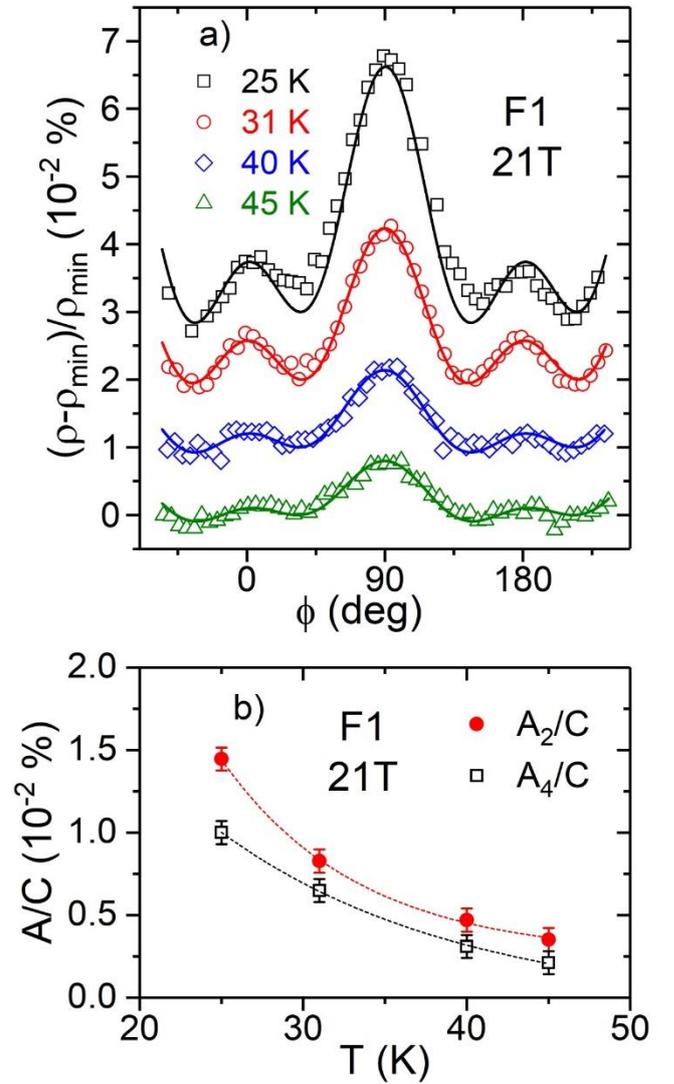

FIG. 5. (a) AMR of film F1 at 21 T and 25 K (black squares, offset by 0.03%), 31 K (red circles, offset by 0.02%), 41 K (blue diamonds, offset by 0.01%) and 45 K (green triangles). Solid lines represent fits to experimental data given by Eq. (1), while $\rho_{min}$ is the resistivity corresponding to the minimum of AMR curves. (b) Temperature dependence of normalized amplitudes $A_2/C$ and $A_4/C$ of twofold (red solid circles) and fourfold (black open squares) oscillations, respectively, at 21 T obtained from (a) by the use of Eq. (1). Error bars are indicated and lines are guides to the eyes.

With the lack of possibility to observe directly the spin configuration (neutron scattering experiments necessitate single crystals), we have to rely on the spin-dependent scattering of the electrons unveiled by the MR of SLCO thin films. Let us consider now the *less doped sample* F2. For $H = 10$ T the AMR is minimum (or the absolute value of MR maximum or the resistivity scattering minimum) whenever $H$ is parallel to $a$-axis ($\phi = 0$, $\pi$, or $2\pi$) and to the current $I$. A second shallower minimum occurs for $H$ along $b$, i. e. perpendicular to $a$ or $I$ ($\phi = \pi/2$). A maximum of $\Delta\rho/\rho(\phi)$ (or a minimum of the absolute value of MR or maximum scattering) occurs for $H$ aligned along the Cu-Cu axis ($\phi = \pi/4, 3\pi/4$). This would suggest that the easy axis is along Cu-O-Cu direction and lead to think that the spin configuration of the undoped SLCO is still preserved (spin along $a$ or $b$ axis). For the highest field, $H = 21$ T, AMR now has a small maximum for $H$ parallel to Cu-O-Cu, a minimum for $H$ along the Cu-Cu axis ($\phi = \pi/4, 3\pi/4$), and the maximum for $H$ along the $b$ axis perpendicular to the current ($\phi = \pi/2$). This suggests that the easy axis is now along the Cu-Cu direction. This change occurs for 16 T $< H <$ 17 T [Fig. 4(f)]. Considering now the *more doped sample* F1, it can be seen that a fourfold AMR is progressively revealed with an increasing magnetic field, with minima of scattering for $H$ along the Cu-Cu direction ($\phi = \pi/4, 3\pi/4$) as if the easy axis was Cu-Cu. As a result, both F1 and F2 exhibit the same AMR behavior at 21 T, and possibly the same final spin configuration. In the more doped sample F1, the spin dilution induced by doping [1] gives way to a disordered spin configuration, not as rigid as in the less doped case, F2. Therefore, the Cu-Cu direction could become, in increasing magnetic field, an easy axis in the more doped sample, while in the less doped sample, spins, presumably first aligned preferably along the $a$ or $b$ axis, would become aligned along Cu-Cu for $H > 16$ T. Finally, a high magnetic field intensity appears to lead to the same final state of the spin configuration in the case of low doping (sample F1) as in the more doped case (sample F2).

### 4. Comparison with the other e-doped family $Ln_{2-x}Ce_xCuO_4$ ($Ln$ = Nd, Pr, La)

The main difference between the $Ln_{2-x}Ce_xCuO_4$ family and the IL SLCO concerns the spin configuration in the AF state. In the $Ln_{2-x}Ce_xCuO_4$ case, the spin configuration, studied by neutron diffraction, is such that in-plane magnetization alternates between adjacent layers in a noncollinear antiferromagnetic structure in the $c$ direction (except for $Ln$ = La with AF collinear structure, see Fig. 16 of [1]). A spin-flop transition, produced by a magnetic field applied in the $CuO_2$ planes, has been reported in these compounds (except $La_{2-x}Ce_xCuO_4$). The spin-flop transition induces a step in the magnetoresistance, $R_{ab}(H)$, and even more specifically in $R_c(H)$, measured with the field $H$ in the $ab$ plane: $H \parallel$ Cu-O-Cu or $H \parallel$ Cu-Cu (see Refs. [8, 12, 37]). Fourfold AMR has also been found in these compounds [11, 38], except in $La_{2-x}Ce_xCuO_4$, where only twofold AMR was observed [13]. In the muon study of $La_{2-x}Ce_xCuO_4$, Saadaoui *et al.* [39] have found that 3D static antiferromagnetism disappears close to the occurrence of superconductivity and that a large extent of the magnetic region, detected by AMR, is due to the fluctuating magnetic moment. In the case of $Ln_{2-x}Ce_xCuO_4$, the rare ion $Ln$ with a magnetic property, especially Nd, may couple with the Cu [1]. Thus Ponomarev *et al.* [40] relate the phase change they observed at 4 T and 1.4 K in the polar representation of the AMR they reported for $Nd_{1.88}Ce_{0.12}CuO_{4+\delta}$ to the combined effect of the spin-flop and field-dependent magnetic Nd contribution to magnetotransport. Even though this phase transition is similar to the one we observe, SLCO does not contain any magnetic ion, and the spin configuration in IL compounds is the same in adjacent $CuO_2$ planes.

Another significant difference between the $Ln_{2-x}Ce_xCuO_4$ family and the IL SLCO is the *much shorter spacing* between $CuO_2$ planes in SLCO that may imply a strong interaction between $CuO_2$ planes (a 3D behavior has been observed for the superconducting properties [30]).

Besides, ARPES measurements on superconducting SLCO thin films ($T_c$ = 25 K) led the authors to conclude that AF and SC coexist simultaneously and homogeneously in SLCO [7]. However, they could not deduce if the AF in SLCO is static or fluctuating.

### V. CONCLUSION

We studied the high magnetic field angular magnetoresistance for two underdoped superconducting SLCO thin films. The films have different doping levels and low-temperature resistance upturns at $T_{min}$. It was previously showed [14] that the more doped sample F1 [$T_c(R = 0) = 17$ K] and the less doped F2 [$T_c(R = 0) < 4$ K] have different AMR behaviors under a magnetic field $H \leq 6$ T: twofold for F1 and fourfold for F2. High magnetic fields (up to 22 T) reveal a negative in-plane magnetoresistance, whose magnitude increases monotonously and linearly and scales with $H/T$. Such behavior is ascribed to the spin scattering. The magnetic field evolution of AMR depends on the sample doping. The more doped sample progressively acquires a fourfold component that becomes visible at $H > 16$ T. This component has a $\pi/4$ phase shift compared with the low-field ($H < 15$ T) AMR of the less doped sample. Unexpectedly, the AMR of the less doped sample F2 undergoes a phase shift of $\pi/4$ at $H = 15$ T with a simultaneous reduction of the amplitude of the fourfold oscillations. With further increase of the field, this amplitude develops again, and both samples possess the *same* fourfold oscillations at high magnetic fields ($H > 16$ T). Such high magnetic field behavior does not change with increasing temperatures ($T < T_{min}$). We tentatively propose that the easy axis for the AF state is the Cu-Cu direction for the more doped sample, while for the less doped one, at fields $H < 15$ T, this axis is in the Cu-O-Cu direction. An intense magnetic field allows the less doped sample to have the same AMR behavior as the more doped one as if the spin configuration was independent of the doping state. Finally, we suggest that AF survives in the normal state of superconducting underdoped SLCO. It would be very interesting to perform muon spin rotation experiments to confirm the presence of magnetism in SLCO.

### ACKNOWLEDGMENTS

V.P.J. acknowledges support from the Ministry of Education, Science and Technological Development of the Republic of Serbia (Contract No. 451-03-68/2020-14/200053). H.R. thanks P. Bourges, M. Gabay, C. Pasquier, P. Senzier, and A. Santander-Cyrot for helpful discussions.

__________________________________________________